\documentclass[trackchanges,twocolumn]{aastex701}

\begin{document}

\title{An unexpected population of quenched galaxies harbouring under-massive SMBHs revealed by tidal disruption events}

\author[orcid=0009-0009-2627-2884]{Paige Ramsden}
\affiliation{School of Physics and Astronomy, University of Birmingham}
\affiliation{Astrophysics Research Centre, School of Mathematics and Physics, Queens University Belfast}
\email[show]{pxr754@alumni.bham.ac.uk}  

\author[orcid=0000-0003-3255-3139]{Sean L. McGee} 
\affiliation{School of Physics and Astronomy, University of Birmingham}
\affiliation{Institute for Gravitational Wave Astronomy, University of Birmingham}
\email{}

\author[orcid=0000-0002-2555-3192]{Matt Nicholl} 
\affiliation{Astrophysics Research Centre, School of Mathematics and Physics, Queens University Belfast}
\email{}

\begin{abstract}

Restricted by event horizon suppression, tidal disruption events (TDEs) provide a unique window into otherwise hidden supermassive black holes (SMBHs) at the lower end of the mass spectrum, allowing the connection between star formation and SMBH mass to be explored across a broad stellar mass range. We derive stellar masses and specific star formation rates using \textsc{Prospector} fits to UV-MIR broadband spectral energy distributions (SEDs) for 42 TDE hosts, together with a high-mass comparison sample, and combine these with SMBH mass estimates from the literature. We first verify our approach by reproducing the established result that quenched galaxies host more massive SMBHs than star-forming systems at fixed stellar mass, a result widely interpreted as evidence for SMBH growth driving the blue-to-red sequence transition. However, examining the TDE sample in isolation reveals a trend reversal at lower masses, uncovering a surprising population of low-mass ($10^{9.6} \lesssim M_{\rm gal} \lesssim 10^{10.5}$\,M$_\odot$), quenched galaxies hosting SMBHs systematically less massive ($M_{\rm BH} \lesssim 10^{6.5}$\,M$_\odot$) than those in star-forming galaxies of comparable stellar mass. After ruling out degeneracies in our SED fits, we conclude that this reflects a physical difference in the quenching mechanism between these TDE hosts and the more massive galaxies. This is unlikely to be driven by AGN feedback, and could instead result from environmental processes, which can end star formation and hinder SMBH growth. We also show that the quenched and post-starburst population within the TDE sample is likely under-represented due to selection biases, suggesting the true fraction could be even higher than observed.

\end{abstract}




\section{Introduction}
\label{s:intro}

Wide-field surveys of the local galaxy population reveal a colour bimodality, with galaxies falling into the blue, star-forming sequence or the red, quiescent sequence \citep{strateva_01, baldry_04}. This well-established divide extends to other galactic properties such as star formation, gas content, morphologies and kinematics \citep{brinchmann_04,catinella_10,cappellari_11}. It is generally understood that low mass galaxies, growing through star formation, become quiescent over time, transitioning from the blue to the red sequence \citep{kauffmann_03a, noeske_07}. More massive galaxies therefore exist in the red regime, growing slowly through mergers \citep{dokkum_05,bell_06}. The transition from blue to red is observed at a characteristic stellar mass of $\sim 10^{10} M_{\odot}$ \citep{kauffman_03b,finlator_08,zinger_20}. The lack of galaxies in the intermediate `green valley' suggests a rapid evolution, rather than a gradual depletion of gas through star formation. While a number of scenarios have been proposed to account for this fast transition, it has so far been difficult to distinguish between them observationally, leaving this stage of evolution poorly understood \citep{pawlik_19}.

Early galaxy evolution models assume that galaxies self-regulate through balancing gas inflows, star formation, and outflows \citep{simon_91,cole_00,finlator_08,bouche_10,dave_12}. Importantly, most massive galaxies host supermassive black holes (SMBHs) at their centres, implying that SMBH feedback should also be considered when treating galaxies as equilibrium states \citep{bower_17, beckmann_17}. Observational evidence for such feedback comes in the form of energetic outflows and radio jets from galactic nuclei \citep{garcia_21,girdhar_22}. Feedback from SMBH accretion can suppress star formation in massive galaxies \citep{bower_17, beckmann_17, zinger_20}, and drive a transition from star-forming to passive. For a given stellar mass, quenched galaxies typically host SMBHs with larger masses than those in star-forming galaxies, as seen in observations, analytic models \citep{bower_17}, and detailed simulations such as EAGLE (Evolution and Assembly of GaLaxies and their Environments; \cite{schaye_15}). However, the observed sample comprises mainly dynamically measured SMBHs, primarily in galaxies with $M_{\rm gal} \gtrsim 10^{10} M_{\odot}$. As a result, the role of black holes in lower-mass galaxies is not well constrained by current data. 

Tidal disruption events (TDEs) occur when an unfortunate star passes too close to a SMBH. At the tidal radius, a star is torn apart by tidal forces, with about half its mass becoming bound \citep{hills_75,lacey_82,rees_88}. If the disruption occurs outside the Schwarzschild radius, a luminous flare is produced. For a main sequence Sun-like star, this happens only below a limiting SMBH mass of $\lesssim 10^8 M_\odot$; above this mass TDEs are unobservable. Restricted by this critical mass, TDEs are successful in identifying low mass SMBHs that would otherwise remain dormant and undetected. 

TDEs also provide insight into unusual host galaxy environments and short-lived phases in galaxy evolution. TDEs are observed  preferentially in E+A or post-starburst (PSB) galaxies, with high central densities, and colours in the `green valley’ \citep{french_16,law-smith_17,graur_18}. Their spectra contain distinct Balmer line absorption and weak, or no, emission lines; indicative of low levels of current star formation, yet significant star formation in the past Gyr such that A-type stars dominate the stellar light. TDEs may therefore provide a unique sample of low-mass SMBHs in green valley galaxies, furthering our understanding of the rapid transition from the blue to the red sequence, as well as shedding light on the connection between SMBH mass and star formation rate at the lower end of the SMBH--stellar mass regime. 

In this work, we model multi-band photometry of 42 TDE host galaxies with \textsc{Prospector}, employing the same model as in our previous work \citep{ramsden_25}. The resulting recent SFHs, galaxy total stellar masses, and measurements of SMBH mass for the TDE sample are combined with a higher mass galaxy sample \citep{kormendy_13}. This joint data set allows us to explore the connection between SMBH growth and specific star formation rate (sSFR) over a large galaxy stellar mass range. Broadly, we find that galaxy evolution is consistent with models dominated by black hole growth. However, in the TDE-only regime, we uncover a reversed trend, revealing an unexpected population of low-mass, quenched galaxies. This surprising result challenges expectations and carries important implications for how we understand this brief phase in a galaxy's lifecycle.

\section{Spectral Energy Distribution Fitting}
\label{seds}

\subsection{Sample Selection and Data}

Our sample of TDE host galaxies is the same as that detailed in \cite{ramsden_25}. We combine the host galaxy total stellar masses and specific star formation rates (sSFRs) obtained through host galaxy modelling, with SMBH mass measurements from \cite{mummery_24}. These SMBH masses are derived from the late-time optical/UV plateauing TDE emission, which is produced by an optically thick accretion disk undergoing expansion and cooling. \cite{mummery_24} show that the plateau luminosity correlates strongly with SMBH mass, a relation that is consistent with both simulated and observed TDEs, providing robust SMBH mass estimates. As a result, the TDE host galaxy sample is restricted to TDEs with plateau detections (49 out of a total of 63 sources). For efficient broadband photometry collection, we use \textsc{Galfetch}\footnote{\url{https://github.com/paigeramsden/GalFetch}}, constructing spectral energy distributions (SEDs) for each host galaxy. All code dependencies and catalogues used are fully described in \cite{ramsden_25}. Availability of broadband photometry imposes further sample cuts; at minimum, multi-colour optical imaging of the galaxy from wide-field surveys such as Pan-STARRS \citep{panstarrs}, SDSS \citep{sdss}, or Legacy \citep{dey_19} is required. After removing candidates lacking sufficient photometry, the final sample comprises 42 TDEs and their host galaxies.

To explore the SMBH--sSFR connection over a broad galaxy stellar mass range the observed TDE sample is combined with a literature sample dominated by higher mass systems. The high mass regime from \cite{kormendy_13} contains SMBH masses measured from stellar, gas and maser dynamics. Following \cite{ramsden_25} we apply several quality cuts, resulting in a final high-mass sample of 42 galaxies. For consistency with the TDE sample, broadband photometry is collected for each of these galaxies and modelled such that total stellar mass is derived in the same way.

\begin{figure*}
\centering
\includegraphics[width=1\textwidth]{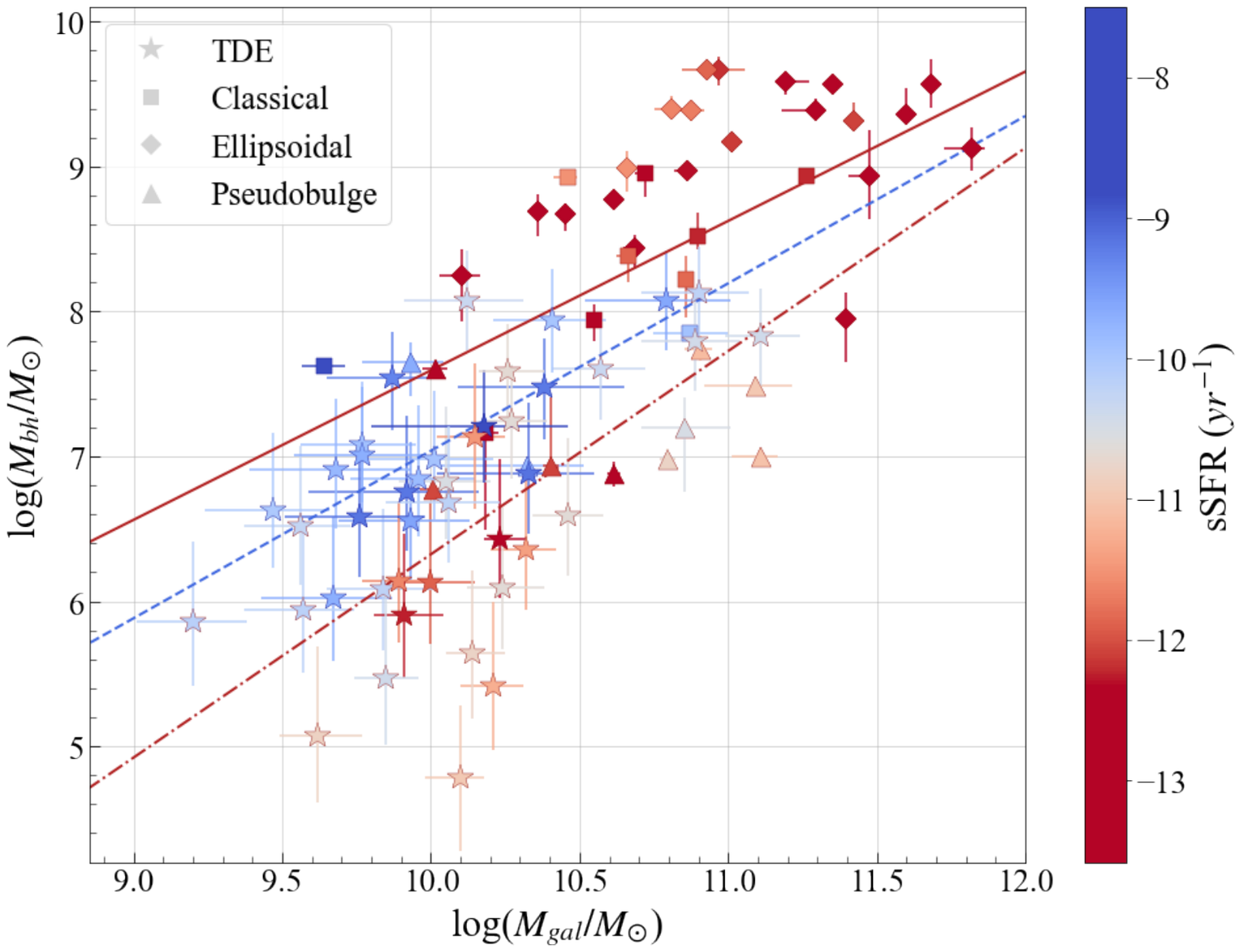}
\caption{SMBH mass as a function of host galaxy total stellar mass for the TDE sample (stars) combined with the high-mass \cite{kormendy_13} sample split into classical, ellipsoidal and pseudobulge galaxy types (squares, diamonds and triangles). Each galaxy is coloured by the logarithm of its specific star formation rate (sSFR). The colour bar is normalised to the range of the TDE sample (10$^{-12}$ $\lesssim$ log sSFR $\lesssim$ 10$^{-9}$). Three lines of best fit computed by least-squares minimisation indicate fits to: high mass regime quenched galaxies (solid, red); TDE star-forming galaxies (dashed, blue); TDE quenched galaxies (dot-dashed, red).}
\label{tde+hmr}
\end{figure*}

\subsection{Prospector Fits}

For modelling the TDE host galaxies, we use \textsc{Prospector} \citep{leja_17, leja_21}, a Python based code that infers stellar population parameters from broadband photometry. In this study, we apply the same model as that used in \cite{ramsden_25}, similar to \textsc{Prospector}-$\alpha$ detailed by \cite{leja_19}. \textsc{Prospector} derives the posterior probability distributions of model free parameters, using dynamic nested sampling with \textsc{dynesty}, including: total stellar mass, metallicity, a six-component non-parametric SFH, and a dust parameter controlling the optical depth from general interstellar dust throughout the galaxy. 

In this work, we target the total stellar mass and recent star formation rate, marginalising over all other parameters. We adopt a broad log-uniform prior on stellar mass, allowing the photometric data to constrain its value based on the SED amplitude.

\textsc{Prospector} offers a number of different SFH priors -- here we use a non-parametric Dirichlet treatment. While non-parametric priors are more computationally expensive, they are less rigid than the parametric alternatives, allowing key properties to vary freely within a set of fixed bins and enabling more flexible modelling of SFHs. The Dirichlet SFH prior was chosen based on the work of \cite{leja_19}, which tested the ability of several non-parametric priors in recovering a range of input SFHs. While all priors performed well, the Dirichlet treatment with a concentration parameter of $\alpha = 0.2$, was found to be the least biased at recovering the widest range of input SFHs, making it well suited to recover the diverse SFHs expected in TDE host galaxies. In this work, we focus on the current star formation rate (SFR) as measured by the final time bin (i.e., the last $\sim 7.5$ Myr). Using a flexible non-parametric prior reduces the influence of the overall SFH shape on the inferred SFR in this final bin, enabling a more robust measurement of recent star formation.

The remaining free parameters used in this model are well described in \cite{leja_17}.

\section{SMBH--sSFR Connection}
\label{smbh-ssfr} 

In Fig.~\ref{tde+hmr} we show the observed relation between SMBH mass and host galaxy total stellar mass from \textsc{Prospector} for the combined TDE and high-mass sample \citep{kormendy_13}. Note that the sample of galaxies plotted here are the same as in \cite{ramsden_25} Appendix A, with the addition of two targets: AT2022wtn and AT2023clx.  Each galaxy is coloured by the logarithm of its specific star formation rate (sSFR). The sSFR is calculated by dividing the star formation rate in the most recent time bin by the galaxy's total stellar mass. Galaxies with a sSFR below $10^{-10.15}$ $\rm yr^{-1}$ would require longer than a Hubble time to double their mass. We therefore adopt this threshold to classify galaxies as passive and plot three lines of best fit computed by least-squares minimisation: high-mass regime passive (solid, red), TDE regime star-forming (dashed, blue), TDE regime passive (dot-dashed, red).

Overall, we see that the more massive galaxy sample from \cite{kormendy_13} is made up primarily of red passive galaxies. In comparison, the TDE sample is lower mass and consists of a mix of star-forming and passive systems. When fitted with the three least-square minimisation lines, clear offsets between star-forming and quenched systems are observed. However, within the TDE host galaxies, passive systems host SMBHs at systematically lower masses compared to star-forming systems of the same stellar mass -- a very surprising result.

\begin{figure}
\centering
\includegraphics[width=1.1\linewidth]{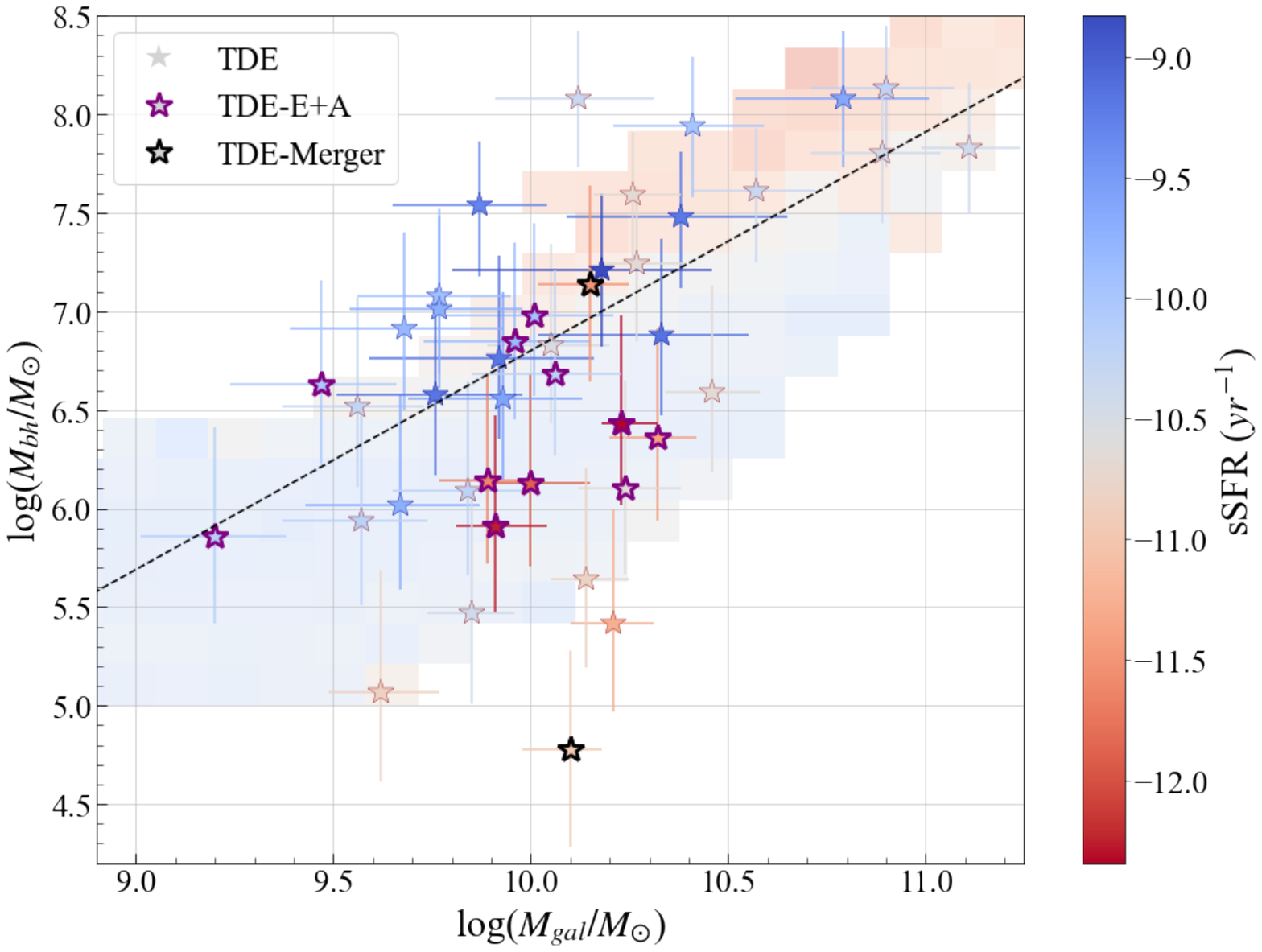}
\caption{SMBH mass as a function of total stellar mass for the TDE host galaxy sample. Each galaxy is coloured by the logarithm of its specific star formation rate (sSFR). The background 2D density map shows galaxies from the EAGLE simulation, coloured using the same sSFR metric. The colour bar is normalised to the range of the TDE sample (10$^{-12}$ $\lesssim \log $(sSFR/yr$^{-1}$) $\lesssim$ 10$^{-9}$). E+A and merging galaxies are highlighted by purple and black outlines respectively. The dashed line (see text for details) derived in \cite{ramsden_25} splits the sample 50:50 -- we find that galaxies below the line are dominated by quenched systems, while those above are primarily star-forming.}
\label{tde}
\end{figure}

Motivated by this finding for TDE hosts, we now focus in more detail on the properties of this sample, selected uniformly by the presence of a TDE. Fig.~\ref{tde} highlights the SMBH–stellar mass relation for TDE host galaxies and allows us to quantify differences between quenched and star-forming hosts. Here we also compare our result to the EAGLE simulation, with the background 2D density map showing bins that contain at least five EAGLE galaxies, coloured using the same sSFR metric. In EAGLE, as in other SMBH feedback models, low-mass galaxies remain blue because their gas supply is regulated by ongoing star formation and stellar-driven outflows. In contrast, higher-mass systems turn red once SMBH-driven feedback becomes dominant, heating, expelling, or cutting off the cold gas supply needed for star formation \citep{white_91,birnboim_03,croton_06,bower_17,byrne_23}. Within the TDE hosts, we identify a distinct population of low-mass quenched galaxies ($10^{9.6} \lesssim M_{\rm gal} \lesssim 10^{10.5} M_{\odot}$), with BH masses $\lesssim 10^{6.5} M_{\odot}$, which is not seen in the simulated EAGLE galaxies. If a scaling relation were fitted exclusively to the quenched TDE hosts, the inferred SMBH--stellar mass relation would be even steeper than that reported in \cite{ramsden_25}. This further emphasises the SMBH mass offset exhibited by quenched TDE hosts.

In order to compare more systematically how the properties and environments of the TDE hosts vary with SMBH mass, we divide our data into two sub-samples. From \cite{ramsden_25}, we take the straight line representing the SMBH–total stellar mass relation of the TDE-only sample and shift it vertically, modifying only the intercept, such that 50\% of the sample sit above the line (i.e.~have a large SMBH mass for their stellar mass) and 50\% sit below (a low SMBH mass for their stellar mass). We can now quantify the differences in sSFR between these two sub-samples. Using the same threshold of sSFR $\lesssim 10^{-10.15}$ $\rm yr^{-1}$ to define passive galaxies, we find that 18/21 galaxies below the 50:50 line are quenched, compared to only 8/21 galaxies above the line. A binomial test shows that both subsamples deviate significantly from the overall quenched fraction of 26/42. The probability of obtaining 18 or more quenched galaxies out of 21 below the 50:50 line is $\sim$1.8\%, while the probability of 8 or fewer quenched galaxies out of 21 above the line is $\sim$2.3\%. This confirms that quenched galaxies are over-represented below the line and under-represented above it. Therefore, the split clearly shows that the star-forming properties of TDE hosts are strongly correlated with SMBH mass, and produces one sub-sample dominated by quenched galaxies (with less massive SMBHs), and another dominated by star-forming galaxies (with more massive SMBHs).

\begin{figure}
\centering
\includegraphics[width=1.05\linewidth]{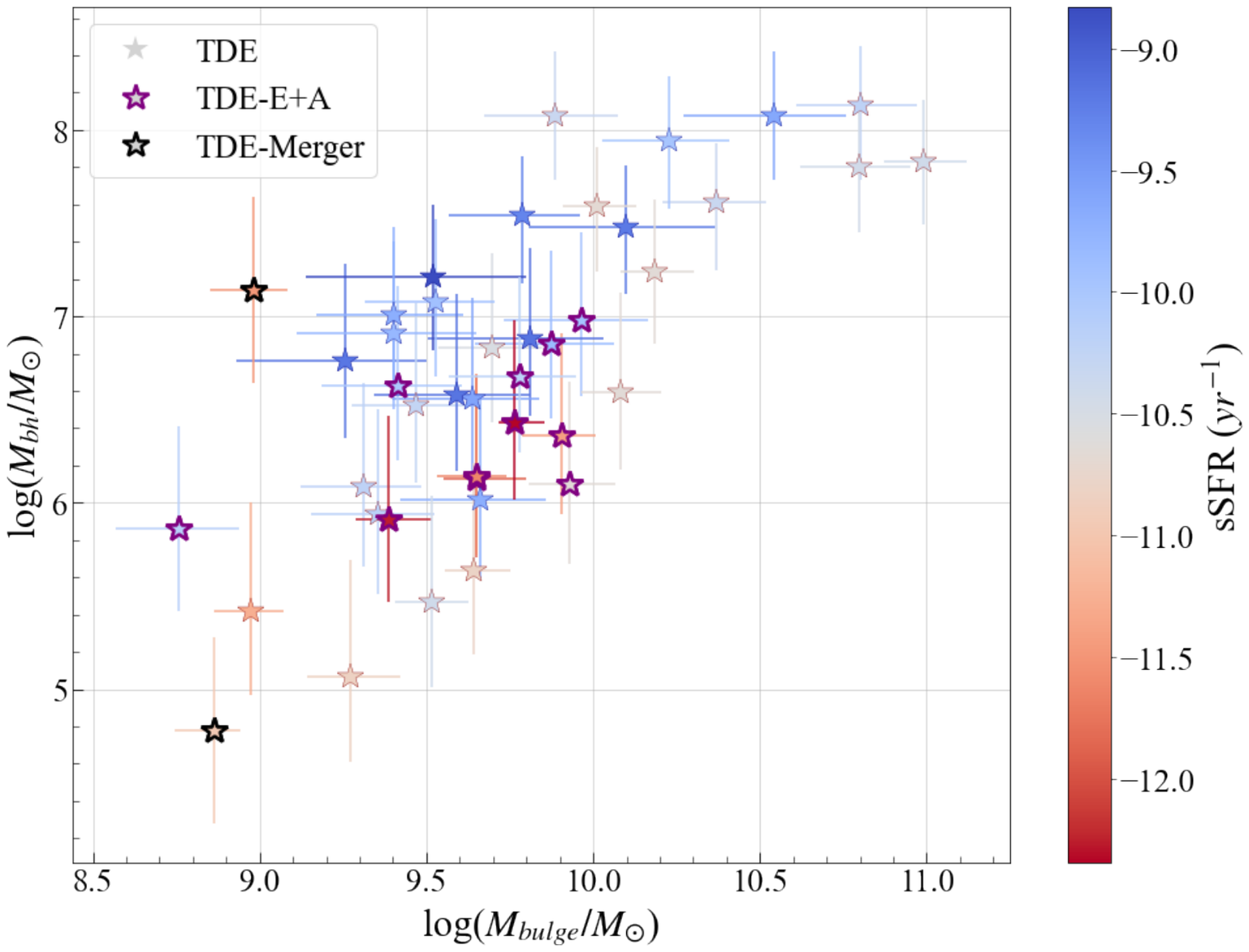}
\includegraphics[width=1.05\linewidth]{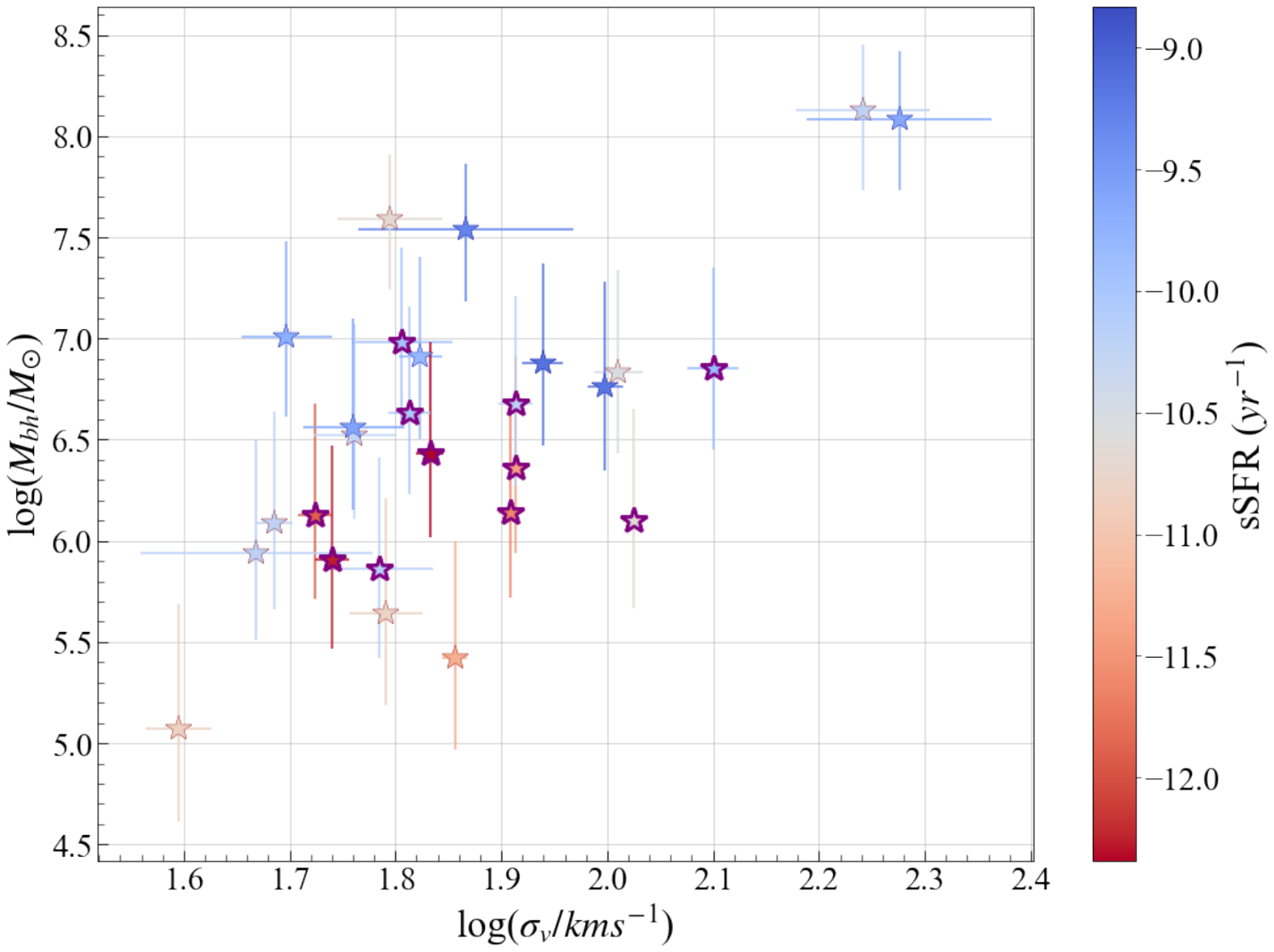}

\caption{Top: SMBH mass as a function of bulge mass for the TDE host galaxy sample. Bottom: SMBH mass as a function of velocity dispersion for the TDE host galaxy sample. Each galaxy is coloured by the logarithm of its specific star formation rate (sSFR). The colour bar is normalised to the range of the TDE sample (10$^{-12}$ $\lesssim \log $(sSFR/yr$^{-1}$) $\lesssim$ 10$^{-9}$). E+A and merging galaxies are highlighted by purple and black outlines respectively.}
\label{tde-bulge-veldisp}
\end{figure}

We also investigate whether the quenched galaxies could appear offset to low BH mass due to differences in their bulges -- i.e., do we observe the same offset in low-sSFR galaxies when plotting BH mass against bulge (rather than galaxy-integrated) properties? In Fig.~\ref{tde-bulge-veldisp}, we show SMBH mass as a function of host galaxy bulge mass (top) and velocity dispersion (bottom), with each galaxy again coloured by $\log(sSFR)$. Here, the velocity dispersion measurements are obtained from the literature \citep{wevers_17, kruhler_18, wevers_20, hammerstein_23, yao_23}, as summarised in \citet{mummery_24}, and bulge masses are taken from \citep{ramsden_25}. It is clear in both cases that the passive population at low SMBH mass is still uncovered. We discuss this further in section \ref{quenching}.

\subsection{Ruling out model degeneracies}

However, before interpreting the observed sSFR trend as a genuine physical difference in TDE host galaxies, we must first consider that the complex nature of modelling broadband galaxy SEDs can lead to significant degeneracies in stellar population fitting \citep{leja_17}. In particular, we aim to rule out known degeneracies between parameters such as dust attenuation, stellar age, and recent star formation. Two especially relevant examples are the dust–recent SFR and age–metallicity degeneracies. Increased dust attenuation can redden a galaxy's SED in a way that mimics the effect of reduced recent star formation, making it difficult to determine whether a galaxy is truly quiescent or simply dust-obscured. Additionally, a young, metal-rich stellar population can produce a similar SED to an older, metal-poor one, complicating the interpretation of derived ages and SFHs \citep{worthey_99}. Furthermore, choices in how to treat the SFH can also influence the inferred present day star formation rate. This latter issue is mitigated by our use of a non-parametric SFH, rather than enforcing any particular functional form to relate past and present star formation. After examining the two-dimensional posteriors for each modelled TDE host galaxy, we find no evidence for degeneracies between the current star formation rate and any of the dust, metallicity, or star formation rate bins.

Additionally, by finding evidence for the expected trend in high-mass systems, we effectively establish a control sample of galaxies fitted in the same \textsc{Prospector} framework as the TDE hosts. If degeneracies were systematically biasing the sSFR result, both samples would be similarly impacted. As this is not the case, the reversal of the trend in the TDE-only regime is unlikely to be driven by systematic modelling effects. Taken together, these results support the interpretation that the sSFR trend in TDE hosts reflects a physical difference in their recent SFHs, rather than a coincidental result driven by degeneracies in the SED modelling. The uncovered population of passive TDE host galaxies may therefore indicate a different quenching mechanism.

\section{Quenching in low-mass TDE hosts}
\label{quenching}

While stellar feedback, such as that from supernovae, is important in regulating the gas content of galaxies, alone it is not energetic enough to quench massive systems, i.e. galaxies around and above the knee of the stellar mass function ($M_{\rm gal} \gtrsim 10^{10.5} M_{\odot}$) \citep{dekel_86, benson_03}. Active galactic nuclei (AGN) can inject energy through jets, radiation, or winds, disrupting the inflow of the cool gas that fuels star formation \citep{springel_05, schawinski_06, schawinski_07}. In theoretical models, AGN feedback energy is often assumed to be a fixed fraction of the SMBH mass \citep{croton_06}. As a result, more massive black holes are linked to more efficient quenching, consistent with observations that high-mass galaxies tend to be red and passive \citep{silk_98, bower_17}. 


However, the systematically low SMBH masses in passive TDE hosts suggests an alternative quenching mechanism in these galaxies. Such a scenario is particularly interesting in the context of TDEs, which are known to occur disproportionately in green-valley and post-starburst galaxies that have been quenched only recently. Understanding what causes the quenching can help to uncover the physical reason for the elevated TDE rate in these systems as well as illuminating the rapid galaxy transition.

For galaxies in groups or clusters, quenching can result from environmental processes such as strangulation, harassment and ram pressure stripping, which can deplete their gas supply and suppress star formation \citep{moore_98, quilis_00, balogh_00, peng_2010}. \cite{geha_12} examine how such environmental factors influence the relative abundance of quenched and star-forming dwarf galaxies. Using a sample sourced from the SDSS Data Release 8 spectroscopic catalogue, they find that quenched dwarf galaxies are preferentially found near massive companion galaxies, with the quenched fraction declining as distance from said companion increases. They also identify a stellar mass threshold of $M_{\rm gal} < 10^9 M_{\odot}$ below which quenched galaxies are not observed in the field, but only in groups -- indicating that galaxies in this mass range are quenched by environmental processes. While this is below the mass regime of our TDE sample, for higher mass galaxies ($10^{9.75} M_{\odot}$) a similar trend is observed: the fraction of quenched galaxies still decreases with distance from the nearest massive companion \citep{geha_12}. Moreover, for field galaxies of stellar mass $\sim 10^{10} M_{\odot}$, the quenched fraction remains low at $\sim 20\%$. Examining the masses of the TDE hosts, it is clear that they occupy an intermediate mass regime between low-mass galaxies, quenched primarily by environmental effects, and high-mass galaxies quenched by AGN feedback. This suggests that environmental effects could be responsible for quenching in TDE hosts. 

In Fig.~\ref{tde+hmr}, we see significant overlap between so-called pseudobulge galaxies and our TDE sample. Pseudobulges follow a tight SMBH–velocity dispersion relation, but with a noticeably shallower slope than that of classical and elliptical bulges, suggesting slower SMBH growth in these systems \citep{hu_08, jiang_11, kormendy_13}.  Additionally, we now see that pseudobulges not only overlap with TDEs in SMBH–stellar mass space, but also in terms of sSFR, with some quenched pseudobulges hosting SMBHs smaller than those in star-forming pseudobulges of similar mass. Although pseudobulges can mimic classical bulges, which are thought to be built via mergers, they are instead thought to form gradually through the secular evolution of gas within the galactic disk \citep{kormendy_n_kennicutt, fisher_08}. Because their SMBHs grow slowly, these passive pseudobulges are unlikely to have been quenched by AGN feedback, supporting the idea that environmental processes could drive quenching in low-mass galaxies. 

Intriguingly, Fig.~\ref{tde-bulge-veldisp} shows that the same correlation with sSFR persists when SMBH mass is plotted against either bulge mass or velocity dispersion. This indicates that, for these low-mass galaxies, quenching is not directly driven by internal properties such as bulge growth or stellar kinematics. Instead, this points to a connection with external factors, motivating an investigation of the role of the galaxies' large-scale environment. This scenario is consistent with \citep{peng_2010} in which mass quenching and environment quenching are separable. Using demographics of galaxies, they find the dominant quenching of low-mass galaxies comes from the environment mode.

Recently, the Euclid Quick Data Release (Q1; \citealt{euclid_25}) investigated the relationship between star formation, stellar mass, and environment across a large sample of galaxies at $0.25 < z < 1$. The study finds that in dense environments, quiescent galaxies are more common at low stellar masses, whereas in the field, star-forming galaxies dominate, particularly at higher masses. This provides further evidence that environmental processes can suppress star formation in low-mass galaxies before internal processes become significant.

We attempted to investigate the environments of our quenched and star-forming TDE hosts by comparing their nearest-neighbour counts. However, at the low redshifts of our sample, galaxies are overwhelmingly dominated by unrelated neighbours that are simply nearby in projection. Spectroscopic redshifts are therefore required for a future experiment.

\subsection{The connection between dense environments and E+A galaxies}

A number of studies have investigated the over-enhancement rates of TDEs in E+A galaxies, also termed post-starburst (PSB) or quiescent Balmer-strong (QBS), finding that a substantial fraction occur in these volumetrically-rare, often quenched, systems \citep{french_16, law-smith_17, graur_18}. The over-representation of TDEs in E+A galaxies has often been attributed to a post-merger origin. Host environments formed through merger may have a high central density of stars \citep{stone_16, law-smith_17}, perturbed central dynamics \citep{stone_18} and/or a central SMBH binary system \citep{mockler_23} -- all of which may enhance the TDE rate. It is thought that the merger triggers a burst of star formation that can rapidly consume available gas and inject energy into the system, preventing further cooling and leading to large-scale shutdown of star formation \citep{matteo_05, hopkins_06}. Since mergers will occur more often in dense environments, a merger origin for the E+A TDE hosts would be consistent with environmental quenching playing an important role for TDE hosts more generally.

We highlight known E+A galaxies within our TDE sample in Fig.~\ref{tde}. Additionally, we mark two TDEs occurring in ongoing mergers, AT2022wtn and AT2023clx \citep{charalampopoulos_24,hammerstein_25} -- in both cases, the TDE occurs in the less massive galaxy of the merging system, with mass ratios of $\sim$10:1. Of the highlighted E+A/merger galaxies, 9/13 sit below our dividing line (i.e.~low SMBH mass for their stellar mass), while the remaining 4 galaxies are positioned just above the line but within 1-$\sigma$. Our E+A/merger sample has an average SMBH mass of $\log(M_{\rm BH}/M_{\odot}) = 6.31 \pm 0.16 M_{\odot}$, compared to $6.71 \pm 0.13$ for the full galaxy sample. This indicates that TDEs identified in E+A/merger hosts have preferentially low SMBH mass, and indeed visually they fall within the mainly quenched region of Fig.~\ref{tde}. We note that 3/13 of highlighted E+A galaxies have sSFR values indicative of active star formation. Nevertheless, it is possible for E+A galaxies to retain a cold gas reservoir during merger that can return them to the blue sequence \citep{french_15, rowlands_15}. 

Any explanation for the E+A enhancement in TDE hosts must account for our observation that they have lower-mass SMBHs at fixed stellar mass. One possible explanation looks to the end result of mergers: a central SMBH binary system. Binary-induced dynamical perturbations can excite the eccentricities of nearby stars, sending them on near radial orbits that intersect one of the SMBH's paths, boosting the TDE rate \citep{li_15, fragione_18, li_19}. SMBH masses that appear systematically smaller than expected for a given galaxy stellar mass could indicate that a TDE has happened around the lower-mass component. However, \cite{mockler_23} find that in cases where the secondary SMBH is an order magnitude less massive than the primary, the resulting number of TDEs around each SMBH can be comparable. We should therefore expect to observe a similar TDE rate from both the primary and secondary SMBH. In other words, if we assume mergers result in E+A galaxies that host central SMBH binaries, around which TDE rates are elevated, we would expect to see a similar fraction of E+A galaxies hosting high and low SMBH masses. Yet, as shown in Fig.~\ref{tde}, we observe an enhancement only at low masses. This disfavours binary SMBHs as the explanation for the low SMBH masses in the E+A galaxies.

Instead, the reason that both quenched galaxies and the E+A subset tend to host lower-mass SMBHs may be due to stunted SMBH growth. Maximal SMBH growth is expected when the halo mass reaches $\sim 10^{12} M_{\odot}$, at which point star formation driven outflows can no longer escape and gas builds up in the central regions \citep{bower_17, byrne_23, scharre_24}. If interactions in a group or cluster environment have already depleted a galaxy’s gas supply, the SMBH is not triggered and both star formation and SMBH growth are suppressed. This naturally produces a `leftover’ population of low-mass, quenched galaxies in which the SMBH--stellar mass correlation persists because growth is limited but not completely halted. Hence, environmental quenching can provide a unified explanation for the uncovered passive population. It can simultaneously inhibit SMBH growth and halt star formation, while the same dense environments also increase merger rates, forming another quenching pathway and producing the observed fraction of E+A galaxies.

\begin{figure}
\centering
\includegraphics[width=1\linewidth]{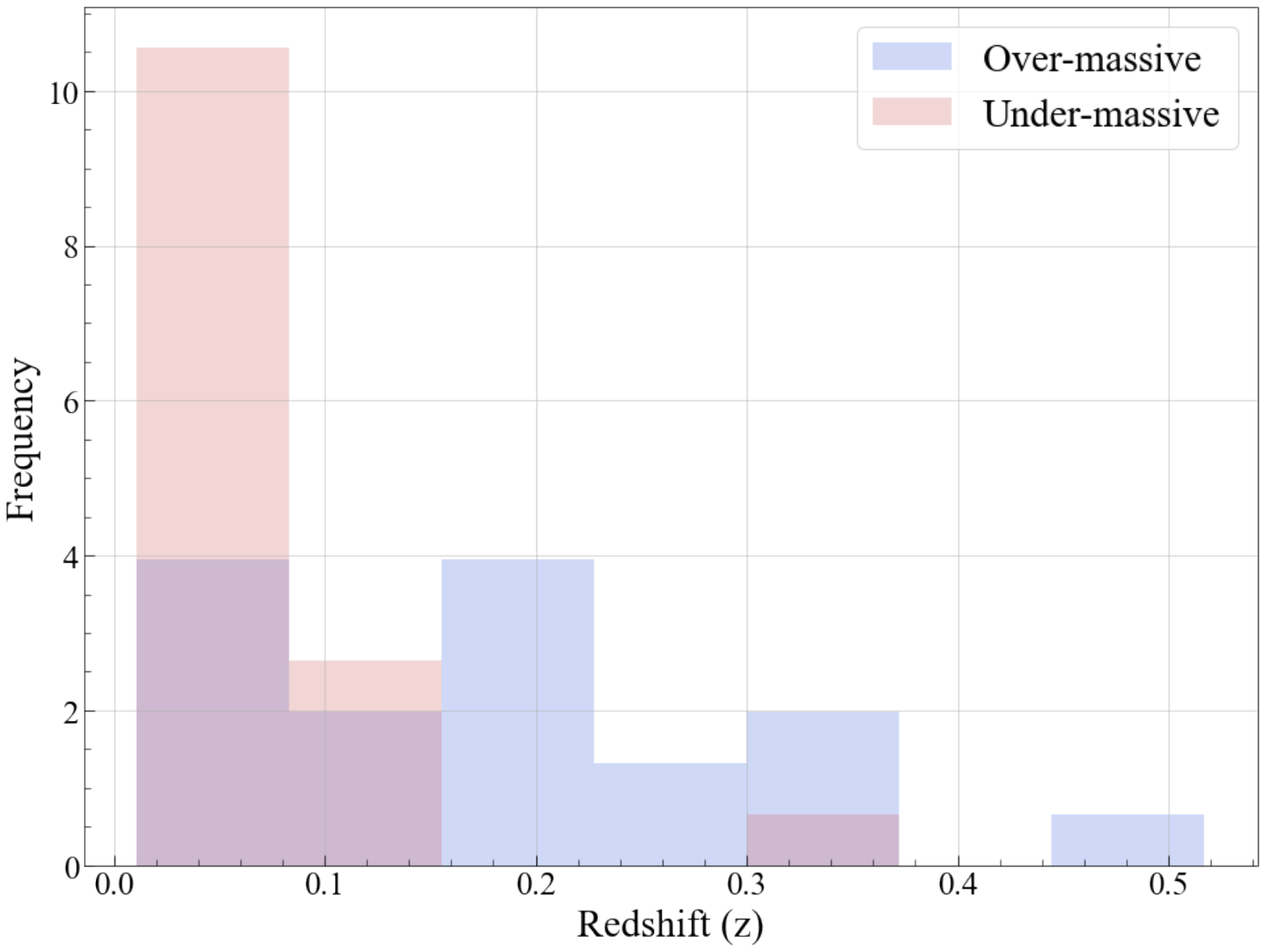}
\caption{Redshift distribution of the two host galaxy sub-samples detailed in section \ref{quenching}. Galaxies below the dashed 50:50 line, hosting under-massive SMBHs and dominated by quenched systems, are shown in red, while those above the line, hosting over-massing SMBHs and typically star-forming, are shown in blue.}
\label{redshift-distribution}
\end{figure}

\section{Redshift biases and the true fraction of E+A hosts}
\label{redshift}

Beyond galaxy types and environments, the redshift distribution of TDE hosts may offer further insight into the recovered low-mass, quenched population. Fig.~\ref{redshift-distribution} shows the redshift distribution of the two host galaxy sub-samples: galaxies below the 50:50 line, hosting under-massive SMBHs and dominated by quenched systems, are shown in red, while those above the line, hosting over-massive SMBHs and typically star-forming, are shown in blue. It is clear that the under-massive SMBH sub-sample is composed of low-redshift galaxies, with a mean value of $z \sim 0.073$. In comparison, the over-massive sub-sample covers a broader redshift range, with a mean value of $z \sim 0.182$.

Over the redshift range probed here ($z \lesssim 0.5$), the ratio of quiescent and star-forming galaxies is not expected to change significantly. Studies of galaxy demographics suggest at most a 10–20\% change by $z \sim 0.5$, with negligible differences below $z \sim 0.2$, where the bulk of our sample is found \citep{weaver_23}. Therefore, rather than reflecting an intrinsic property of TDE hosts, the apparent skew towards low redshift in the population of (mostly quenched) galaxies with under-massive SMBHs may instead result from selection biases that limit the detectability of relatively low mass TDEs at higher redshift. \cite{mummery_24} find that the TDE peak luminosity is proportional to SMBH mass. Looking at the two sub-samples defined earlier, the over-massive SMBH sub-sample dominated by star-forming galaxies has an average SMBH mass of $\log(M_{\rm BH}) = 7.29 \pm 0.50$, whereas the under-massive sub-sample, made up largely of quenched galaxies, has an average of $\log(M_{\rm BH}) = 6.13 \pm 0.66 M_{\odot}$. This implies that TDEs in the over-massive, typically star-forming sample are $\sim 14\times$ brighter on average than those in the under-massive, often quenched sample. Consequently, TDEs occurring in star-forming host galaxies around more massive SMBHs can be detected at distances up to $\sim 3.7\times$ greater than those in red, passive galaxies. If we examine only galaxies at $z<0.1$, the ratio of quenched to passive galaxies is 17:6, compared to 13:8 for the overall sample, and the fraction of E+A galaxies is 9/13. This may be more representative of the volume-corrected quenched/E+A overabundance.

\section{Conclusions}

To gain insight into the physical factors impacting galaxy quenching and the TDE rate, we have investigated the relation between central SMBH mass and star formation rate in TDE host galaxies. We modelled the spectral energy distributions of 42 TDE host galaxies alongside a sample of high-mass galaxies from \cite{kormendy_13} with \textsc{Prospector}.

Our main conclusions are as follows:

\begin{itemize}

    \item In the TDE sample alone, we uncover a distinct population of low-mass quenched galaxies, $10^{9.6} M_{\odot} \lesssim M_{\rm gal} \lesssim 10^{10.5} M_{\odot}$. These galaxies host SMBHs with masses $\lesssim 10^{6.5} M_{\odot}$, too low to drive large-scale quenching through AGN feedback, and inconsistent with a simple monotonic relationship between SMBH mass and quenching.

    \item After ruling out known degeneracies, we determine that the sSFR trend in TDE hosts reflects a genuine physical difference indicative of a different quenching mechanism within this population.

    \item We divide the TDE only sample along a shifted SMBH--stellar mass line such that 50\% falls above the line and 50\% below. This creates a sample of over-massive and under-massive SMBHs, $\log(M_{\rm BH}/M_{\odot}) \sim {7.29 \pm 0.50}$ and ${6.13 \pm 0.66}$ respectively. We find that the under-massive sample is dominated by quenched galaxies, while the over-massive sample typically contains star-forming galaxies. 
    
    \item We find that 9/13 of known E+A/merging galaxies fall into the under-massive SMBH sub-sample dominated by passive galaxies. The remaining 4/13 sit within $\sim$ 1-$\sigma$ of the cut between the two sub-samples. The high fraction of quenched E+A/merging galaxies aligns with a post-merger scenario for E+A galaxies, which predicts that mergers are responsible for their lack of star formation.

    \item Environmental quenching offers a unified explanation for the uncovered passive TDE hosts, simultaneously stunting SMBH growth, halting star formation, and boosting merger rates to produce the observed E+A fraction.

    \item The more passive TDE sub-sample is composed of low-redshift galaxies, with an average of $z \sim 0.073$ compared to the more active sub-sample which has an average of $z \sim 0.183$. This apparent skew of quenched systems towards low redshift implies the presence of selection biases. In a volume-limited TDE sample, the fraction of low-mass passive systems and E+A galaxies could in fact be much higher than currently observed.

\end{itemize}

Constrained by the critical Hills mass, TDEs enable the study of otherwise hidden low-mass SMBHs, revealing a distinct population of passive galaxies that do not fit into the standard picture of galaxy quenching. Looking forward, targeted spectroscopic follow-up is required to characterise the environments of these systems and understand the transition from the blue to red sequence in the lower-mass regime. Furthermore, the Rubin Observatory will be instrumental in tackling the apparent over-representation of these passive systems at low redshift \citep{lsst, bianco}. With its unprecedented survey depth, Rubin galaxy photometry will soften the current magnitude limit and reduce biases that favour the detection of low-mass star-forming galaxies. This will enable either the discovery of a larger population of such passive galaxies at higher redshifts or confirmation of their restriction to low redshifts.

\section*{Acknowledgements}

PR acknowledges support from STFC grant 2742655.

MN is supported by the European Research Council (ERC) under the European Union’s Horizon 2020 research and innovation programme (grant agreement No.~948381).

\section*{Data Availability}

This article is based on existing publicly available data. \textsc{GalFetch} queries: SDSS \citep{sdss}, PanSTARRS \citep{panstarrs} and DECAM \citep{decam} for optical data in $u$, $g$, $r$, $i$, $z$, $y$\footnote{$u$ bands from SDSS (and DECAM where possible), $y$ bands from PanSTARRS (and DECAM where possible).}; 2MASS, for near infrared (NIR; \cite{2mass}); unWISE, for mid-infrared (MIR; \cite{unwise}); GALEX, for UV \citep{galex}.

\software{Astropy \citep{astropy_13,astropy_18}, Astroquery \citep{astroquery_19}, Extinction \citep{extinction_21}, Extinctions \citep{extinctions_git}, Mechanize \citep{mechanize_git}, Numpy \citep{numpy_20}, Pandas (\hspace{-0.25em}\citet{pandas_20}, Requests \citep{requests_git}}

\bibliography{refs}{}
\bibliographystyle{aasjournalv7}



\end{document}